\documentclass[manuscript]{aastex}
\def\H2{\hbox{H$_{2}$}}

\def\one{{\,\sc i}}
\def\two{{\,\sc ii}}

\def\four{{\,\sc iv}}

\begin{document}

\title{Background Infrared Sources for Studying\\ the Galactic Center's Interstellar Gas}

\shortauthors{Geballe et al.}

\shorttitle{Background IR Sources in the Galactic center}

\author{T. R. Geballe\altaffilmark{1}}
\author{E. Lambrides\altaffilmark{2}}
\author{B. Schlegelmilch\altaffilmark{3,4}}
\author{S.  C. C. Yeh\altaffilmark{5}}
\author{M. Goto\altaffilmark{6}}
\author{Calvin Westrick\altaffilmark{7,8}}
\author{T. Oka\altaffilmark{7}}
\author{F. Najarro\altaffilmark{9}}

\altaffiltext{1}{Gemini Observatory, 670 N. A'ohoku Place, Hilo, HI 96720; tgeballe@gemini.edu}
\altaffiltext{2}{Department of Physics \& Astronomy, Johns Hopkins University, Bloomberg Center, 3400 N. Charles St., Baltimore, MD 21218, USA}
\altaffiltext{3}{Department of Physics and Astronomy, University of California, Los Angeles, California 90095, USA}
\altaffiltext{4}{current address: School of Engineering, Massachusetts Institute of Technology, 77 Massachusetts Ave., Cambridge, MA 02139}
\altaffiltext{5}{W. M. Keck Observatory, 65-1120 Mamalahoa Hwy., Kamuela, HI 96743, USA}
\altaffiltext{6}{Universit\"ats-Sternwarte M\"unchen, Ludwig-Maximilians-Universit\"at, Scheinerstrasse 1, 81679 M\"unchen, Germany}
\altaffiltext{7}{The Enrico Fermi Institute, University of Chicago, Department of Astronomy and Astrophysics and Department of Chemistry, Chicago, IL 60637 USA}
\altaffiltext{8}{current address: Department of Mathematics, University of Minnesota, Minneapolis, MN 55455}
\altaffiltext{9}{Centro de Astrobiologõ«a (CSIC-INTA), Ctra. Torrejo«n a Ajalvir km 4, 28850 Torrejo«n de Ardoz, Spain}

\begin{abstract}

We briefly describe the results of a K-band spectroscopic survey of over 500 highly reddened point-like objects on sightlines toward the Central Molecular Zone (CMZ) of the Galaxy. The goal was to find stars with featureless or nearly featureless spectra suitable for near- and mid-infrared absorption spectroscopy of the Galactic center's interstellar gas on sightlines spread across the CMZ. Until recently only a few such stars have been known outside of very localized sightlines in the vicinity of the Quintuplet and Central clusters. We have used {\it Spitzer Space Telescope} (GLIMPSE) and 2MASS photometry to  select promising candidates, and over the last ten years have been acquiring low-resolution K-band spectra of them. As expected, the vast majority are cool and/or highly reddened red giants with complex photospheric spectra unsuitable for measuring faint interstellar lines. Approximately ten percent of them, whose observations are reported here, have featureless or nearly featureless spectra.  Although not evenly distributed in Galactic longitude, these stars are scattered across the CMZ.  Many of them are luminous stars that are deeply embedded in warm dust cocoons, and have K-band continua rising steeply to longer wavelengths. A significant fraction of them are hot stars of a variety of spectral types, including at least five newly discovered Wolf-Rayet stars.   All of them should be suitable for spectroscopy of interstellar absorption lines at infrared wavelengths greater than 3~$\mu$m and many are also suitable at shorter wavelengths.  

\end{abstract}


\section{INTRODUCTION}

Until recently, it was generally thought that the interstellar gas within the central few hundred parsecs of the Galaxy, a region of angular dimensions approximately $2.2^{\circ} \times 0.4^{\circ}$ usually referred to as the Central Molecular Zone (CMZ), consists of two major environments: ultra-high temperature (10$^{6-8}$ K) plasma responsible for the apparently diffuse X-ray emission and enhanced scattering of radiation from background radio sources, and molecular clouds with densities $\gtrsim$ 10$^{3}$ cm$^{-3}$ \citep{koy89,yam90,mor96, laz98, oka98}. In the last fifteen years, high-resolution absorption spectroscopy of H$_3^+$ at 3.5--4.0 $\mu$m and CO at 2.3 $\mu$m and 4.7 $\mu$m has revealed the presence of another gaseous environment in the CMZ. This gas, with densities $\lesssim$10$^2$ cm$^{-3}$, has the characteristics of Galactic diffuse cloud material, but is considerably warmer (200--300 K), and appears to be expanding radially from a location near the center \citep{got02, oka05}.  Its density, temperature, and velocity are derived virtually entirely from spectroscopy of three lines of H$_3^+$. Most of this warm diffuse gas is not physically associated with the CMZ's dense clouds \citep{oka05}. The measured column densities of H$_3^+$  toward objects in the Central Cluster and in and near the Quintuplet Cluster ($\sim$30 pc distant from Sgr A* on the plane of the sky) indicate absorption path lengths for it that are significant fractions of the dimensions of the CMZ, and thus suggest that within the CMZ it has a large and possibly dominant volume filling factor \citep{oka05}. Its significant filling factor must come mainly at the expense of the large filling factor of the ultra-high temperature gas proposed earlier \citep[see the discussion in][]{oka19}.

To test this interpretation of the motion and distribution of the CMZ's warm diffuse gas, absorption spectra of H$_3^+$ are needed on widely spaced sightlines across the CMZ. Until recently, however, only a handful of stars, all located in or close to the aforementioned two clusters, were known to have sufficiently featureless spectra (so that the faint interstellar lines are uncontaminated by photospheric absorption lines) and to be sufficiently bright for high-resolution spectroscopy at 3.5--4.0~$\mu$m on 8m class telescopes. This paper describes a program that has addressed that need. 

\begin{table}[http]
\caption{Telescopes and Spectrographs }
\label{tels}
\begin{center}
\begin{tabular}{cccc}
\hline
Telescope & Spectrograph & slit width & R  \\
\hline
UKIRT & UIST & 0\farcs24 & 1000 \\
Gemini N & NIRI & 0\farcs48 & 780 \\
Gemini N & GNIRS & 0\farcs45 & 1200 \\
Gemini S & FLAMINGOS-2 & 0\farcs54 & 600 \\
IRTF & SPEX & 0\farcs80 & 750 \\
\hline
\end{tabular}
\end{center}
\end{table}

\section{OBSERVATIONS AND DATA REDUCTION}

Over the last ten years we have been conducting a K-band spectroscopic survey of bright and highly reddened infrared stars located on sightlines to the CMZ to search for ones with suitably smooth spectra that can be used for velocity-resolved spectroscopy of lines of H$_3^+$, which have optical depths of at most 0.1 and are often much weaker. The survey has used five infrared spectrographs at four telescopes: the 1--5-$\mu$m Imaging Spectrograph, UIST, at the United Kingdom Infrared Telescope (UKIRT), the Near Infrared Imager (NIRI) and the Gemini Near-Infrared Spectrograph (GNIRS) at the Federick C. Gillett Gemini North Telescope,  SPEX, at the NASA Infrared Telescope Facility (IRTF), and FLAMINGOS-2, at the Gemini South Telescope. The observations were made using a variety of slit widths (0\farcs24--0\farcs80) and at a variety of resolving powers, summarized in Table 1. Weather conditions varied considerably on many of the nights when survey observations were undertaken.  Most of the spectra were obtained at Gemini North and Gemini South in poor weather conditions that were unusable for almost all other science programs.  Limited amounts of time specifically for this program were allocated at UKIRT, Gemini North, and the IRTF.  A few of the observations were made at telescopes on nights when it was impossible to carry out other science programs that had been scheduled. 

The survey used photometry from the Two-Micron All-Sky Survey \citep[2MASS;][]{skr06} and  the {\it Spitzer Space Telescope} Galactic Legacy Infrared Midplane Survey Extraordinaire \citep[GLIMPSE;][]{ram08} to identify candidate stars. (In this paper we refer to these stars by their 2MASS right ascensions; e.g., 2MASS J17425693-2940428 is referred to as 17425693.) The selection criteria evolved somewhat during the ten-year survey, but always strictly required that the candidates be on sightlines toward the CMZ (Galactic longitude within 1.2$^{\circ}$ and latitude within 0.3$^{\circ}$ of Sgr A* and have {\it Spitzer} IRAC [3.6~$\mu$m] $ <$ 8~mag. Additional less rigorous criteria included: (1) IRAC [3.6~$\mu$m] $-$ [7.9~$\mu$m] $\gtrsim$ 1.5 mag; (2) 2MASS $J$ $\gtrsim$ 12 mag; and (3) 2MASS $J - K$ $\gtrsim$ 4 mag. Selected targets in crowded fields also needed to be clearly identifiable in 2MASS images in order to unambigiously locate them when performing spectroscopy. Candidates located in the Central Cluster within about one arc-minute of Sgr A* were excluded. 

Over 500 candidate stars were identified as satisfying the strict criteria and at least some of the additional three criteria. Due to the difficulty in separating the effects of extinction to the Galactic center (typically 30 visual magnitudes, but with considerable spatial variation), extinction physically associated with the candidates themselves, and possible thermal emission from warm dust shells associated with candidates, it was impossible to determine based on photometry alone which candidates would be usable as probes of the interstellar gas. Low-resolution K-band spectra can readily determine suitability, however, because cool stars, such as red giants, which have complex atomic and molecular absorption spectra rendering them unsuitable, have prominent CO absorption bands at 2.3--2.4~$\mu$m that are easy to detect in quick low-resolution spectra, even in poor weather conditions.  As expected, the vast majority ($\sim$90 percent) of candidate stars turned out to be red giants.  Their spectra will be published elsewhere (Geballe et al., in prep.).

Data reduction utilized both custom pipelines and manual reduction with existing IRAF and Figaro tools. In all cases, standard techniques of spike removal, flatfielding, rectification, extraction, wavelength calbration, and ratioing by early-type telluric standards were employed. Depending on the seeing, the extracted spectra covered 0\farcs5--1\farcs5 along the slit of the spectrograph.  Telluric standards were usually, but not always, observed on the same nights as the candidate objects.  Flux calibration was not attempted due to the variable sky conditions, but the data reduction preserves the slopes of the K-band continua of the candidate stars.

\begin{table}[http]
\renewcommand{\arraystretch}{.8}
\caption{Background Sources}
\label{sources}
\begin{center}
\tiny
\begin{tabular}{ccccccccccccc}
\hline
2MASS ID&{\it l}$^{II}$&{\it b}$^{II}$&Obs.Date&Tel/Instr&J$^a$&H$^a$&K$^a$&[3.6]$^b$&[7.9]$^b$&Type$^c$&Other Names / New$^d$&Prev. K Sp.\\
\hline
17425693-2940428&359.063&0.108&20140412&GN/GNIRS&17.00&3.32&10.66&7.48&5.48&RG&New ($\beta$+-)& \\ 
17431001-2951460&358.931&-0.029&20080818&UKIRT/UIST&17.01&13.91&11.58&7.02&4.13&D&New($\alpha$-)& \\
17432173-2951430&358.953&-0.065&20080728&UKIRT/UIST&14.03&9.37&6.48&3.79&1.72&D& New ($\alpha$)&e\\ 
17432823-2952159&358.958&-0.090&20100521&GN/NIRI&17.62&12.81&10.10&7.02&4.94&RG&New ($\alpha$+)& \\
17432988-2950074&358.992&-0.076&20080731&UKIRT/UIST&17.54&12.56&8.82&4.53&1.01&D&New($\beta$)& \\
17443734-2927557&359.435&-0.091&20110412&GN/GNIRS&17.56&13.91&10.29&7.03&5.48&WR&New($\gamma$-)& \\ 
17444083-2926550&359.456&-0.092&20080731&UKIRT/UIST&16.76&12.81&9.40&6.61&4.58&WR&New ($\gamma$)& \\
17444319-2937526&359.305&-0.195&20150518&GS/F2&16.00&12.78&10.14&7.47&5.56&OB&New ($\beta$++) &\\
17444501-2919307&359.569&-0.041&20100521&GN/NIRI&14.37&10.97&9.07&7.04& 3.48&misc&New($\gamma$++)& \\ 
17444840-2902163&359.820&0.099&20140408&IRTF/SPEX&13.81&12.08&10.34&7.70&5.46&misc&New ($\delta$++)& \\
17445461-2852042&359.977&0.168&20120501&GN/GNIRS&15.79&13.87&11.28&6.64&3.46&D&New ($\epsilon$-+)& \\ 
17445538-2941284&359.277&-0.264&20150511&GN/GNIRS&16.72&12.63&10.14&7.63&6.02&OB&New ($\beta$+) &\\
17445945-2905258&359.797&0.037&20160929&GS/F2&8.72&7.59&7.023&6.55&5.99&OB&G359.717+0.037&f \\
17450241-2854392&359.955&0.121&20140805&IRTF/SPEX&15.81&12.56&10.00&6.66&4.30&OB&New& \\
17450483-2911464&359.717&-0.035&20080815&UKIRT/UIST&15.02&11.63&9.04&6.51&4.87&WR & New($\delta$)& \\
17450929-2908164&359.775&-0.018&20100517&IRTF/SPEX&15.15&13.17&11.19&7.65&5.04&OB&Pa3-141 ($\delta$+)&g \\
17451618-2903156&359.859&0.004&...&...&11.49&9.17&7.89&6.76&5.76&misc&X174516.1&g\\
17451917-2903220&359.863&-0.006&...&...&15.51&12.52&9.86&7.64&6.89&WR&X174519.1&g\\
17452405-2900589&359.907&-0.001&20100602&GN/NIRI&13.03&10.34&8.78&6.93&5.23&misc&G359.907-0.001($\epsilon$-)&f\\
17452861-2856049&359.985&0.027&20080815&UKIRT/UIST&14.34&11.26&9.22&6.58&1.43&OB&H2 ($\epsilon$)&h\\ 
17453782-2857161&359.985&-0.011&20110716&GN/GNIRS&15.56&12.30&10.30&7.79&5.10&OB&H8 ($\epsilon$'')&h\\
17454390-2825200&0.451&0.247&20150407&GS/F2&17.67&14.18&10.15&6.47&4.14&RG&New& \\
17455154-2900231&359.967&-0.081&20100521&GN/NIRI&17.61&15.09&11.20&7.56&3.62&OB&D ($\epsilon$')&h\\
17455583-2845189&0.190&0.036&20080818&UKIRT/UIST&14.87&13.40&11.28&7.16&3.98&RG&New& \\
17455585-2837456&0.297&0.102&20120428&GN/GNIRS&16.28&11.61&8.76&6.55&3.70&RG&New ($\theta$+-)& \\
17460164-2855154&0.059&-0.068&20110515&GN/GNIRS&13.34&10.42&9.00&7.62&5.57&OB&Pa3-103 ($\epsilon$++) &i\\
17460215-2857235&0.030&-0.088&20080716&UKIRT/UIST&14.87&11.43&8.08&4.20&0.97&D&New ($\epsilon$+)& \\
17460433-2852492&0.099&-0.055&...&...&14.26&10.19&7.56&4.60&...&...&NHS21, qF577&j\\
17460562-2851319&0.120&-0.048&20150826&GS/F2&12.53&9.24&7.46&6.37&4.51&misc&NHS22, G0.120-0.048&k\\
17460825-2849545&0.148&-0.043&...&...&15.10&10.94&8.33&6.61&444&...&NHS42, q578&j\\
17461292-2849001&0.170&-0.049&20100516&IRTF/SPEX&13.79&10.53&8.85&7.60&5.50&misc&New ($\eta$+)& \\
17461412-2849366&0.163&-0.058&20080728&UKIRT/UIST&15.41&10.88&7.72&3.90&1.00&WR&GCS3-4, Q1&l\\
17461431-2849317&0.165&-0.058&20080728&UKIRT/UIST&15.83&12.47&8.91&4.52&0.80&WR&GCS3-3, Q9&l,m\\
17461471-2849409&0.164&-0.060&20080728&UKIRT/UIST&13.55&10.69&7.29&3.16&0.32&WR&GCS3-2, Q2 &l\\
17461481-2849343&0.165&-0.060&20080728&UKIRT/UIST&13.27&10.46&7.52&4.69&2.31&WR&GCS 3-1, Q4&l\\
17461514-2849323&0.167&-0.060&20080731&UKIRT/UIST&13.29&10.12&8.41&5.57&3.55&misc&Q13 ($\zeta$)&m\\
17461524-2850035&0.159&0.065&...&...&11.83&8.92&7.29&5.63&4.69&misc&NHS25,Pistol&n\\
17461539-2849348&0.166&-0.062&20150614&GS/F2&15.05&11.47&9.10&6.77&5.23&WR&Q6&m\\
17461586-2849456&0.164&-0.065&20080716&UKIRT/UIST&14.90&10.41&7.24&3.53&0.98&WR&WR102ha, GCS4, Q3&l\\
17461783-2850074&0.163&-0.074&20080731&UKIRT/UIST&13.99&10.37&7.84&5.52&3.65&WR&New ($\eta$)& \\
17461798-2849034&0.179&-0.065&...&...&12.31&8.97&7.09&6.41&4.98&misc&FMM362&o\\
17462830-2839205&0.337&-0.013&20100516&IRTF/SPEX&16.88&13.15&10.62&7.52&3.45&OB&X174628.2 ($\theta$+)&g\\
17463219-2844546&0.264&-0.074&20080728&UKIRT/UIST&17.14&12.86&9.21&6.38&3.83&WR&New ($\theta$)& \\ 
17463270-2816282&0.671&0.171&20120528&GN/GNIRS&17.31&14.45&11.34&7.75&4.76&RG&New ($\kappa$+)& \\
17464524-2815476&0.704&0.137&...&...&15.39&9.96&7.18&3.00&0.63&OB&WR&g\\
17470137-2813006&0.775&0.111&20150511&GN/GNIRS& 16.65&13.13&9.98&7.06&5.11&RG&New ($\lambda$-+)& \\
17470898-2829561&0.548&-0.059&20080815&UKIRT/UIST&17.18&15.35&10.45&6.56&...&RG&($\iota$)&e\\
17470921-2846161&0.315&-0.201&20150421&GS/F2&14.21 &12.58&10.18&7.29&...&misc&New ($\theta$+-+)& \\
17470940-2849235&0.271&-0.228&20130622&GN/GNIRS&12.66&10.80&9.46&7.75&5.65&OB&New ($\theta$+--)& \\
17473680-2816005&0.799&-0.026&20100512&GN/NIRI&15.22&14.48&12.41&7.36&4.58&RG&New ($\lambda$)& \\
17474486-2826365&0.663&-0.143&20080731&UKIRT/UIST&17.59&14.22&10.28&4.70&1.49&RG&New ($\kappa$)& \\
17482472-2824313&0.769&-0.250&20120425&GN/GNIRS&15.86&12.12&9.54&6.72&4.09&OB&New ($\lambda$-)& \\
\hline
\end{tabular}
\end{center}

\end{table}

\newpage

\scriptsize
\noindent $^a$2MASS magnitude, except where noted\\
$^b$GLIMPSE magnitude \\
$^c$D = dust, RG =  (veiled) red giant, OB = OB star, WR = Wolf-Rayet star\\
$^d$Greek letters and signs refer to designation of \citet{oka19}\\
$^e$\citet{geb10}\\
$^f$\citet{mau10c}\\
$^g$\citet{mau10a}\\
$^h$\citet{cot99}\\
$^i$\citet{don11} (name refers to table number and source number)\\
$^j$\citet{fig99}\\
$^k$\citet{mau10b}\\
$^l$\citet{naj17}\\
$^m$\citet{lie09}\\
$^n$\citet{fig98}\\
$^o$\citet{geb00}\\

\normalsize

\section{RESULTS AND DISCUSSION}

\begin{figure}  
\centerline{
\resizebox{1.0\textwidth}{!}{\includegraphics[angle=270]{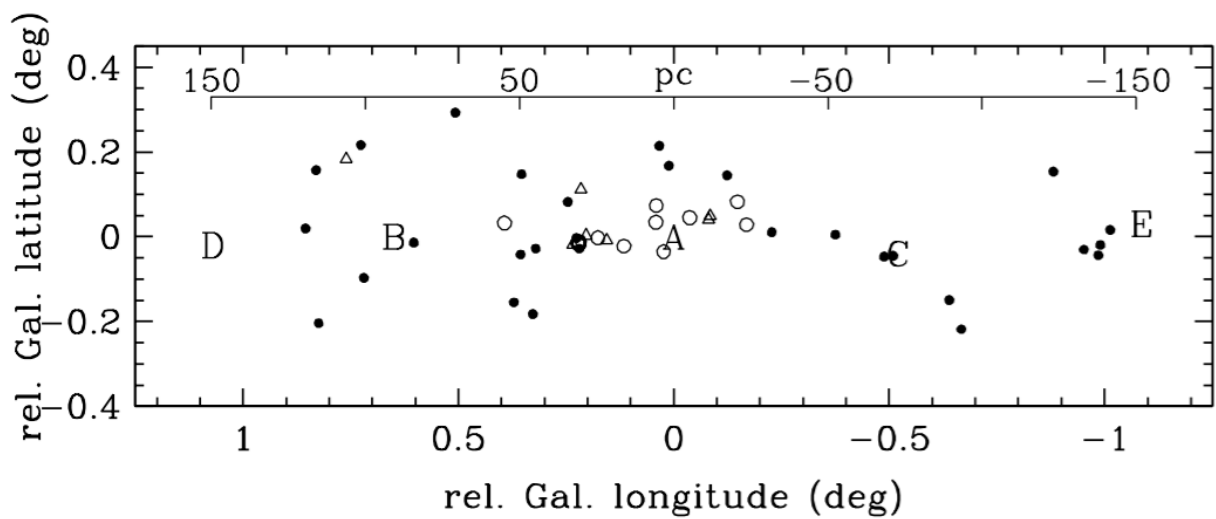}}}
\caption{Locations of the stars listed in Table 1, in Galactic coordinates relative to Sgr A* ({\it l}$^{II}$ = -0.056, {\it b}$^{II}$ =  -0.046 (2000)).  Filled circles are newly found suitable stars; open circles are previously known suitable stars observed in this survey; triangles are suitable stars not observed in this survey. The distance scale is for a galactocentric distance  of 8.0 kpc. Letters indicate locations of the well-known Sagittarius radio sources, Sgr A, B, C, D, E.}
\label{longlat}
\end{figure}

\begin{figure}  
\begin{center}
\includegraphics[width=0.95\textwidth]{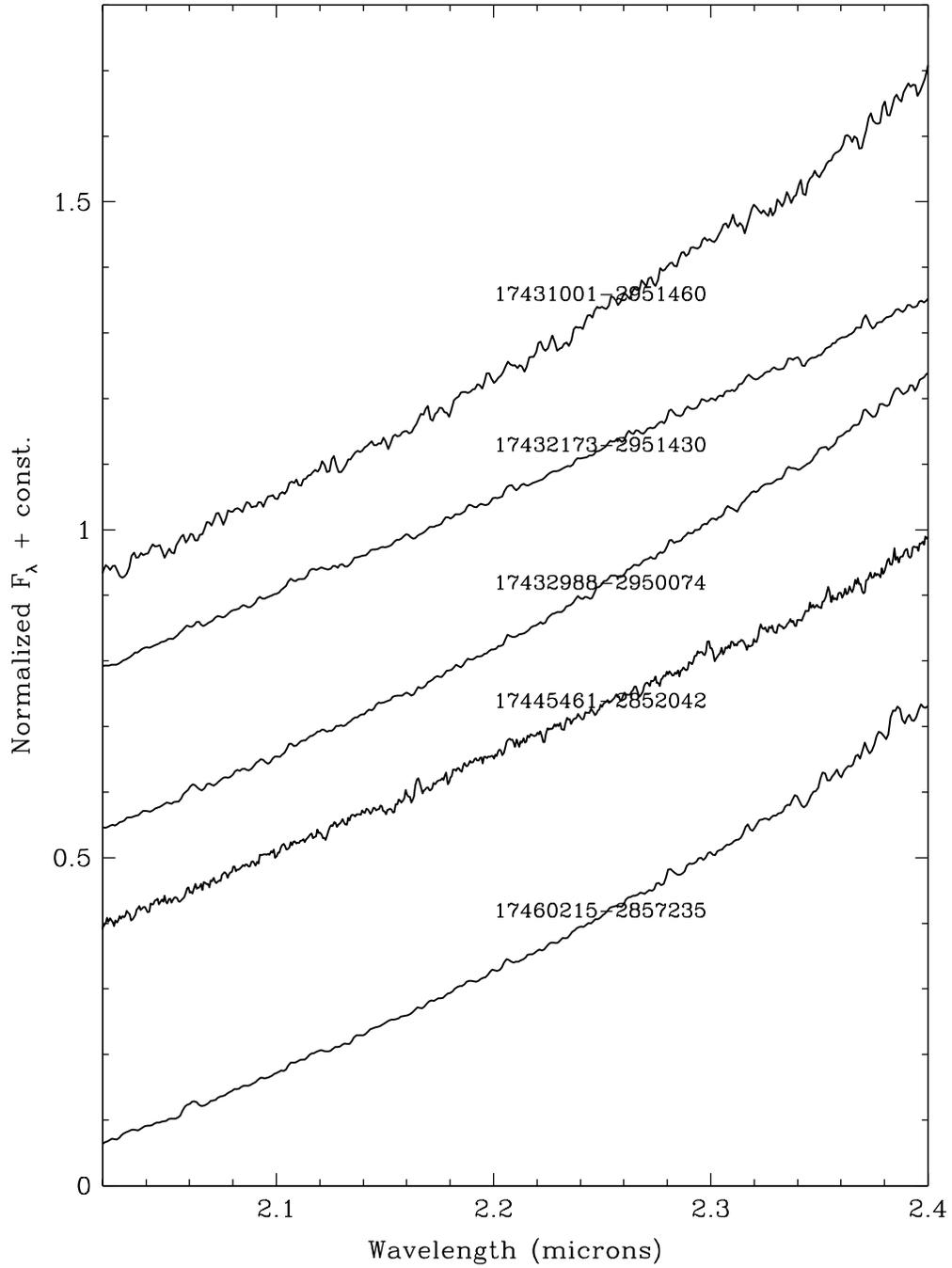}
\caption{Featureless K-band spectra of five stars on sightlines toward the CMZ, with 2MASS identifications. Flux densities are normalized to unity at 2.40~$\mu$m.}
\label{dusty}
\end{center}
\end{figure}

\begin{figure}
\begin{center}
\includegraphics[width=0.95\textwidth]{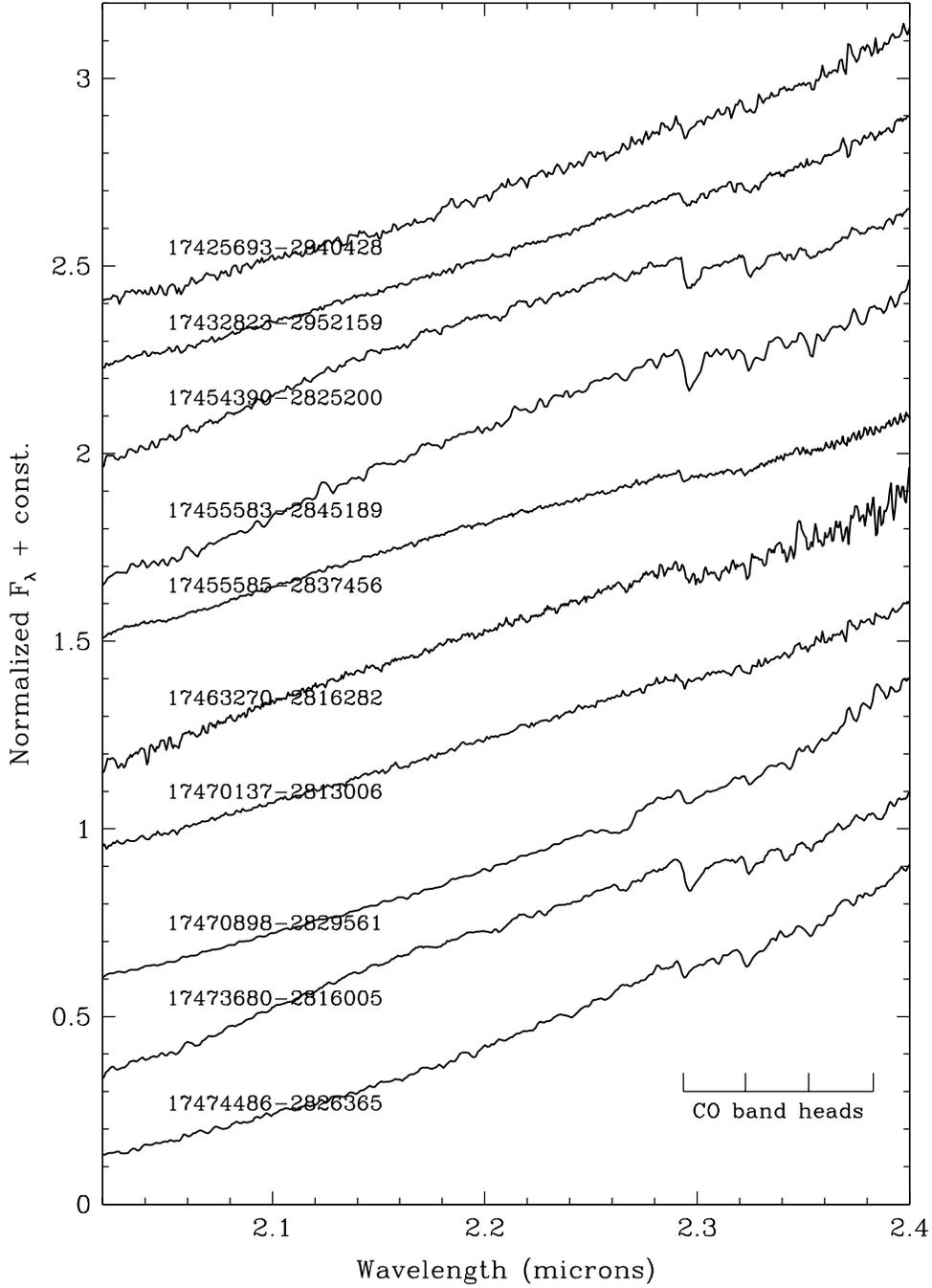}
\caption{K-band spectra of ten stars on sightlines to the CMZ showing weak CO bands and steeply rising continua. The 
wavelengths of the band heads of $^{12}$C$^{16}$O are indicated.} 
\label{veiled}
\end{center}
\end{figure}

\begin{figure}
\begin{center}
\includegraphics[width=0.95\textwidth]{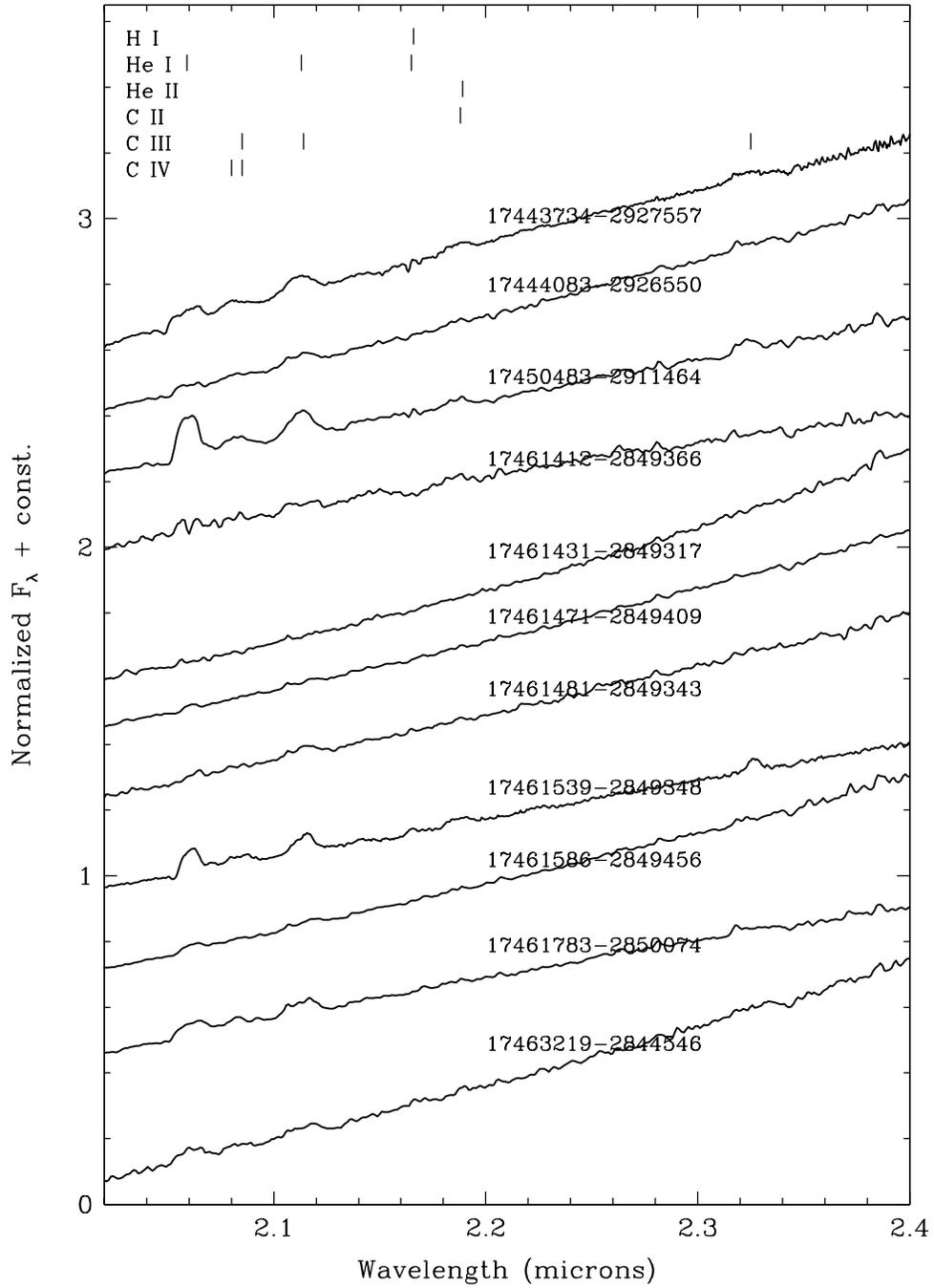}
\caption{K-band spectra of eleven stars on sightlines to the CMZ suspected to be or known to be Wolf-Rayet stars. Line wavelengths are shown at top.}
\label{wr}
\end{center}
\end{figure}

\begin{figure}
\begin{center}
\includegraphics[width=0.95\textwidth]{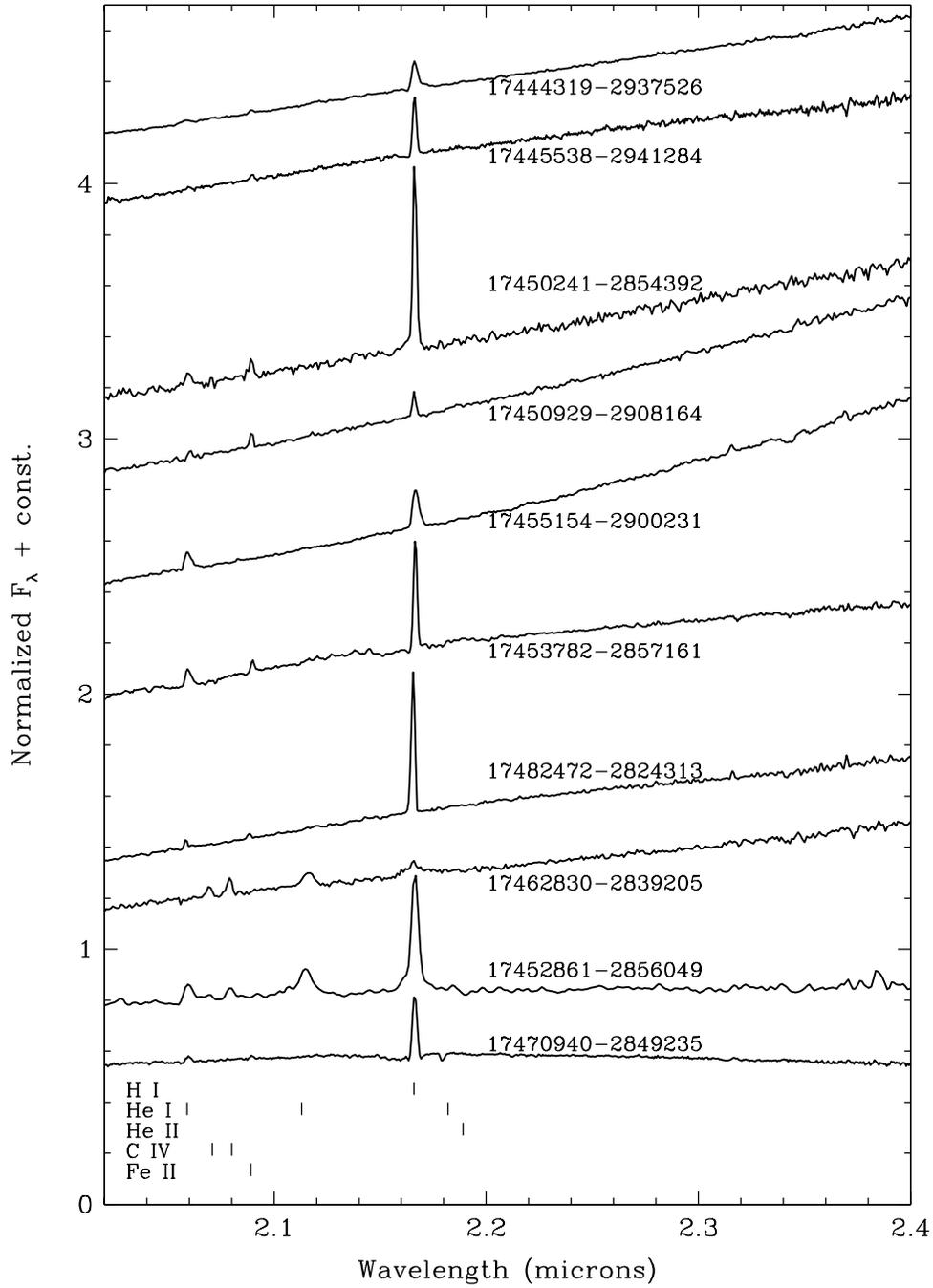}
\caption{K-band spectra of ten stars on sightlines to the CMZ with narrow H I Br $\gamma$ emission and He\one\ emission lines, with line wavelengths shown at bottom.}
\label{hii}
\end{center}
\end{figure}

\begin{figure}
\centering
\begin{center}
\includegraphics[width=0.95\textwidth]{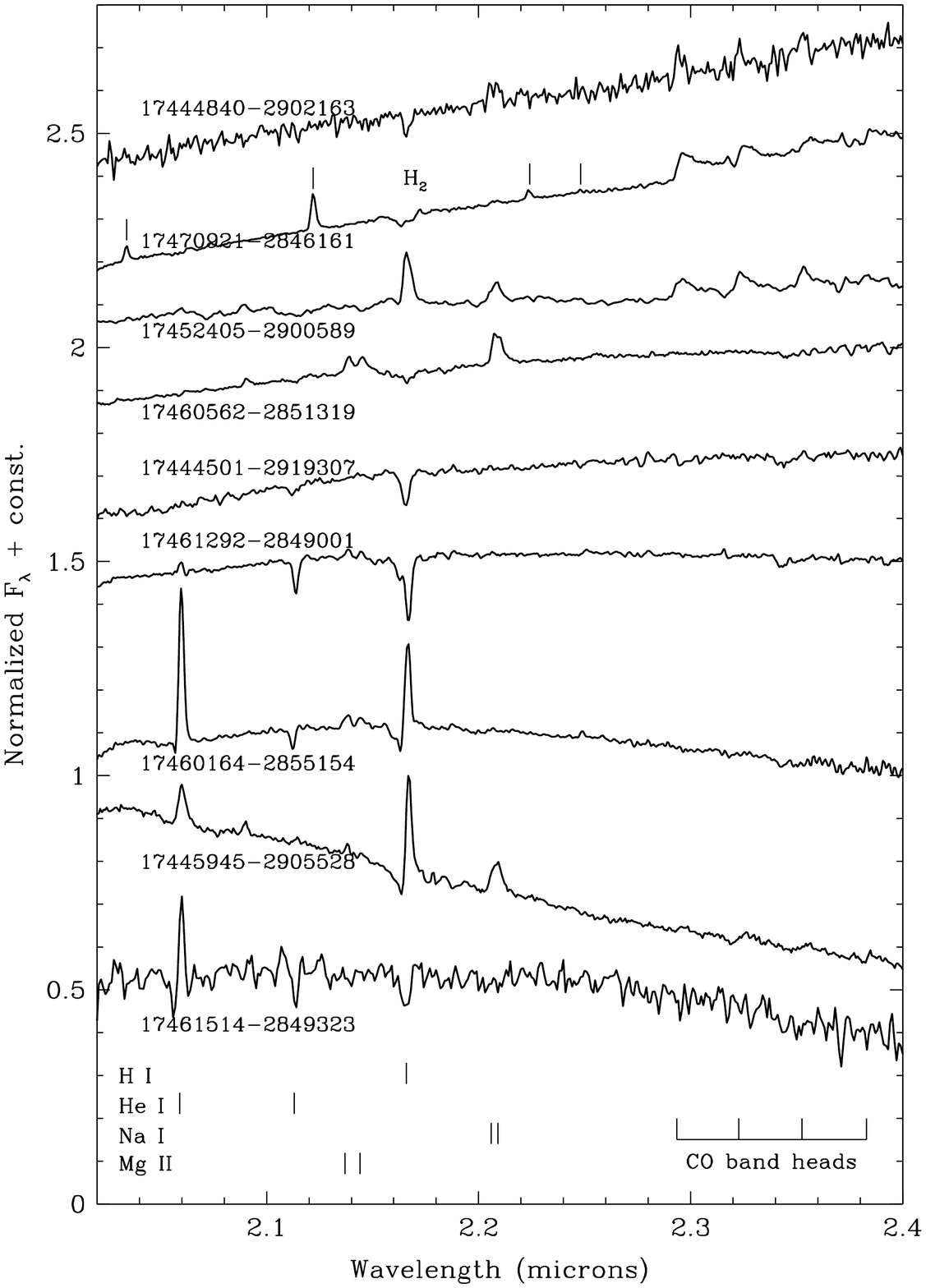}
\caption{K-band spectra of nine stars with diverse atomic and molecular emission and absorption features. Line wavelengths are shown at bottom. For 17470921 wavelengths of H$_{2}$ lines are indicated separately.}
\label{misc}
\end{center}
\end{figure}

The positions of the surveyed stars whose spectra indicate that they will be useful for high-resolution spectroscopy of interstellar absorption lines at $\lambda$ $>$ 3~$\mu$m  are shown in Figure 1. They are listed in Table~2, in order of right ascension, along with some of their relevant properties, including Galactic coordinates.  Their spectra are presented in Figures 2-6.  Figure 1 and Table 2 also include seven suitable stars from the literature, which satisfy the strict criteria, but were not included in the survey. 

In Figures 2-6 each spectrum is scaled so that its flux density is unity at 2.40~$\mu$m. The spectra are grouped into five categories, each of which corresponds to one of the figures. The large positive slopes of most of the continua displayed in Figures 2-6 reflect the high reddening and/or the presence of warm dust cocoons, as discussed in Section 3.1. Sixty percent of the suitable stars in the survey are newly discovered; most of these are located more than 30 pc from Sgr A*. In contrast,  all but two of the stars for which previous K-band spectra exist are located within 30 pc of Sgr A*. 

As can be seen in Figure 1, the distribution of the stars is somewhat clumpy, with concentrations at the location of the Quintuplet Cluster ($\sim$ 30 pc east of Sgr A*) and near the radio source Sgr E. There is an absence of stars in the vicinity of the eastern edge of the CMZ  (near Sgr D). Apart from these attributes, the longitudinal coverage across the CMZ is fairly uniform.  It is known from spectroscopy of lines of H$_3^+$ that most of these stars lie within the CMZ \citep{oka19}. In many cases the low-resolution spectra presented here show a weak absorption feature centered near 2.345 $\mu$m, which is due to cold CO in the ground vibational and low ($J \leq\ 3$) rotational levels. This in itself demonstrates that many are more distant than at least some of the intervening spiral arms, whose interstellar CO produces this absorption, as can be seen in the higher resolution spectra of \citet{oka05} and \citet{oka19}. In the other spectra presented here, low signal-to-noise ratios and/or contamination by overlapping spectral features precludes detection of this absorption. For many of these stars, especially the newly discovered ones, whether and how deeply they are within the CMZ cannot be determined without high-resolution spectroscopy of interstellar lines of species such as H$_3^+$ and CO, as demonstrated in the above two cited papers. In almost all cases where such spectra have been obtained it is clear that the stars are located within the CMZ \citep{oka19}.

\subsection{Stars with Featureless Spectra}

The stars in Figure 2 appear to be completely featureless and heavily obscured by their own dust shells, which are also emitting continuum radiation. Both the obscuraton and dust emission hide the stellar photospheres from view, as argued below. The natures of the embedded stars are unknown. Two of the stars, 17432988 and 17460215, show somewhat broadened emission features at the wavelength of the He\one\  2~$^1$$P$~--~2~$^1$$S$ 2.059~$\mu$m line. This wavelength also corresponds to a strong telluric absorption band of CO$_2$, which suggests that the features could be residuals of the ratioing of the candidates' spectra by those of the telluric standards. In the absence of any other spectral features, we tentatively consider these spectra as featureless, rather than, e.g., characteristic of Wolf-Rayet stars (Section 3.3).

The flux densities of the sources in Figure 2 increase by factors of 2-5 between 2.02~$\mu$m and 2.40~$\mu$m. For a naked star suffering 30 magnitudes of extinction (a typical extinction to stars in the Galactic center), the reddening between 2.02~$\mu$m and 2.40~$\mu$m is 0.77 mag (a factor of 2.03), assuming an extinction law with a  $\lambda^{-1.7}$ wavelength dependence \citep{ind05}.  For an unobscured star whose  K-band continuum is on the Rayleigh-Jeans tail of the blackbody function, the  flux density, $F_\lambda$, at 2.40~$\mu$m is 0.50 that at 2.02~$\mu$m.  The two effects almost exactly offset one another, and thus typically the continuum flux density in the K band of a star in the CMZ that is devoid of circumstellar dust will be flat.  If the observed increases in  F$_\lambda$ with wavelength in Figure 2 were entirely due to cold foreground dust or cold circumstellar dust, it would imply additional  2.02--2.40-$\mu$m reddenings of 0.75--1.75 mag and total visual and extinctions at $V$ and $K$ of  60--100 mag and 5--8 mag, respectively.   The latter are unreasonably large in view of the apparent $K$ magnitudes of the stars (Table 2) and their distances. A similar conclusion is reached if other reddening  estimates with steeper indices but similar values of $A_K$ are used \citep[e.g., $A_{Ks}$  = 2.42, $A_{\lambda} \propto \lambda^{-2.11}$,][]{fri11}. Thus, for many of the stars the principal cause of the rapid increase in flux density with wavelength in the $K$ band must be thermal emission from warm circumstellar dust. This argument applies to most of the stars whose spectra are shown in Figures 3 and 4 and some of the stars whose spectra are shown in Figures 5 and 6. 
 
 \subsection{Stars with Veiled Photospheric Absorption Spectra}
 
The spectra of the stars in Figure 3 contain weak overtone CO band absorption with the $2-0$, $3-1$, and $4-2$ bands clearly visible in most cases. The associated ro-vibrational energy levels of CO can only be populated in high density gas at temperatures of a few thousands of Kelvins, and therefore the bands must arise in the photospheres of the stars. Although it is possible that these photospheric CO bands are intrinsically weak, in view of the rapidly rising continua (comparable to the spectra in Figure 2) and the arguments in the previous paragraph, it is much more likely that their photospheres are ``veiled" by emission from warm dust. Another possibility, which is difficult to exclude in view of the crowded fields and intermittent poor seeing, is that some of these spectra are superpositions of spectra of stars with pure dust continua and red giants. Regardless, at wavelengths $>$ 3~$\mu$m the veiling should be much greater and the stars should be entirely suitable for observing interstellar absorption lines.  

\subsection{Dusty Wolf-Rayet Stars and Candidates}

Nearly all of the spectra shown in Figure 4 contain broad emission lines.  Six of the stars whose spectra are shown, including the two that appear featureless (17461412 and 17461431, also known as GCS 3-4 and GCS 3-3) are already known to be or believed to be dusty Wolf-Rayet stars \citep{tut06,naj17}.  Five of them, including those two, are the well-known Quintuplet stars \citep{gla90,nag90,oku90}. The other five stars whose spectra are shown in Figure 4 are newly discovered. Based on the similarities of their spectra to those of the infrared Quintuplet \citep{naj17}, we conclude that these five also are dusty Wolf-Rayet stars. Of them, two are located within the Quintuplet Cluster. The other three, 17443734, 17444083, and 17450483, are far from that cluster and also are not associated with the other two massive star clusters in the CMZ. 

These new detections suggest that a significant fraction of dusty Wolf-Rayet stars were not detected by the Hubble Space Telescope Pa~$\alpha$ survey \citep{don11}. The latter three, as well as many of the hot and luminous stars whose spectra are shown in Figures 5 and 6, increase the number of apparently isolated massive stars in this region. Their existence raises a fundamental question: are these Òmassive field starsÓ the result of tidal interactions between clusters, escapees from one or more of the existing clusters due to internal events, or representatives of a new mode of massive star formation in isolation. 

\subsection{Emission line stars}

All of the stars in Figure 5 show the H\one\ Br $\gamma$ line in emission and in all but one the He\one\ 2.059~$\mu$m line is  also in emission.  These lines are considerably narrower than the lines in the Wolf-Rayet stars in Figure 4 and thus the stars in Figure 5 are likely to be OB stars.  Because of the limited wavelength coverage of the spectra, we do not attempt to classify them. \citet{mau10a} has classified two of the stars, 17452861 and 17462830 as O4--6I. Three in this group  (17452861, 17453782, 17462830), by virtue of the broad wings or P Cygni profiles on their Br~$\gamma$ (2.166~$\mu$m) and/or He\one\  lines, possess strong high velocity winds.  The Fe\two\ line at 2.089~$\mu$m is in emission in six stars (17445538, 17450241, 17450929, 17470921, 17453782, and 17482472), the C\four\ lines at 2.071 and 2.080~$\mu$m are present in two (17462830 and 17452861), and He\two\ 2.189~$\mu$m is present (weakly) in 17452861, in concordance with the classification of this star by \citet{mau10a}.  

\subsection{Other Stars}

The diverse spectra that are shown in Figure 6 do not appear to fit into any of the categories of spectra shown in Figures 2--5. 
Some of the stars are probably B supergiants or hypergiants.  Although \citet{lie09} assigned a O6-8If spectral type to the Quintuplet star 17461514, \citet{cla18} have  recently reclassified it as B2-3I+. Our rather poor quality spectrum of this object, showing He\one\ 2.059~$\mu$m in emission and  He\one\ 4--3 2.113~$\mu$m and Br $\gamma$ in absorption, appears nearly identical to the one in \citet{lie09}. The spectrum of another likely early B hypergiant  in Figure 6, 17461292, also located in the Quintuplet Cluster, is similar in its He\one\ and H\one\ line profiles, but also has the Mg\two\ doublet at 2.137 and  2.144~$\mu$m in emission.  The spectrum of a third star, 17460164, closely resembles the B1Ia+ star BP Cru (Clark et al. 2018). A fourth star, 17444501 (far from the Quintuplet Cluster) shows He\one\ 4-3 and Br $\gamma$ in absorption, the former indicative of a B2-3 hypergiant. 

The spectra of four of the stars in Figure 6 (17444840, 17445945, 17452405, and 17470921) contain emission features from the  blended Na I doublet (2.206, 2.209 $\mu$m) and from the CO overtone vibrational bands at 2.3--2.4~$\mu$m. Four emission lines of H$_2$, at 2.034, 2.122, 2.224, and 2.248~$\mu$m, the first three from the $v$ = 1--0 band and the fourth (very weak) line from the $v$ = 2--1 band are also present in 17470921. The  CO band emission, which extends at least up to the $v$ = 5 level in all four of these stars, must arise in hot ($\sim$3,000 K) and dense ($n$ $\gtrsim$ 10$^{10}$ cm$^{-3}$) circumstellar material either in the form of disks or winds in close proximity to the stellar photospheres. In the case of 17445945, the P Cygni profiles of the Br $\gamma$ line and the He~\one\ 2.059~$\mu$m line indicate the presence of a strong ionized wind.  The Br $\gamma$ line in 17452405 is in emission, whereas in 17444840 and 17470921 it is in absorption. 

Finally, the spectrum of 17460562, obtained in 2015, differs noticeably from an earlier one reported by \citet{mau10b}, who classified this star as a luminous blue variable. While Fe\two\ 2.089~$\mu$m, the Mg\two\ doublet  and the Na\one\ doublet remain in emission in 2015, the Br $\gamma$ line, previously strongly in emission, is in absorption and the He\one\ 2.059~$\mu$m line, previously in emission, is absent. 

\section{CONCLUDING REMARKS}

The stars whose spectra are reported here cover a wide of Galactic longitudes, extending from three-fourths of the way to the CMZ's eastern boundary all the way to its western boundary.  Many of them have already been used as probes of motion and distribution of the CMZ's warm diffuse interstellar gas \citep{geb10,got11,oka19} at wavelengths between 2.3~$\mu$m and 3.7~$\mu$m. A few have been used to study diffuse interstellar bands in the $J$ and $H$ bands \citep{geb11}. We expect that they will be a valuable resource for future studies of interstellar lines in the CMZ as well as in the foreground spiral arms, including studies at much longer infrared wavelengths.

The list of suitable stars in Table 2 should not be considered as strictly complete to {\it Spitzer} IRAC [3.6~$\mu$m] $ <$ 8~mag.  Seven of the stars in that table were either missed or intentionally not observed (2MASS J17451618-2903156, J17451917-2903220, J17460825-2849545, J17461524-2850035, J17461798-2849034, J17464524-2815476, and NHS21).  In particular, it is possible that some hot stars that are not embedded in warm circumstellar dust and thus do not have very red infrared colors also were not included in the survey.

The limiting $L$ magnitude of 8 was selected in order to ensure that high signal-to-noise ratio measurements of the weak lines of H$_3^+$ in the 3.5--4.0-$\mu$m region could be obtained with the current generation of high-resolution infrared spectrographs on ground-based 8--m class telescopes in integration times of no more than a few hours.  Clearly, a deeper survey than the one reported here should uncover many additional sources, which will be suitable for high-resolution spectroscopy with the next generations of spectrographs and large telescopes. Such a survey is likely to provide a denser and more uniform longitudinal sampling of sightlines across the CMZ. In addition, while it is clear that the current set of background sources extends to the center of the CMZ and possibly in few cases somewhat beyond that \citep{oka19}, a deeper survey would surely find additional suitable background sources that are more deeply within or perhaps even behind the CMZ, allowing gas in its rear half to be studied via absorption spectroscopy. 

\begin{acknowledgements}
 
This research is based in part on observations obtained at the Gemini Observatory (Programs  GS-2003A-Q-33, GS-2008A-C-2, GS-2009A-C-6, GN-2010A-Q-92, GN-2011A-Q-105, GN-2011B-Q-12,  GN-2011B-Q-90, GN-2012A-Q-75, GN-2012A-Q-121, GN-2013A-Q-114, GN-2014A-Q-108, GS-2014A-Q-95,  GN-2015A-Q-402, GS-2015A-Q-96, GN-2016A-Q-96, GS-2016A-Q-102, GS-2017A-Q-95), which is operated by the Association of Universities for Research in Astronomy, Inc., under a cooperative agreement with the NSF on behalf of the Gemini partnership: the National Science Foundation (United States), the National Research Council (Canada), CONICYT (Chile), the Australian Research Council (Australia), Minist\'{e}rio da Ci\^{e}ncia, Tecnologia e Inova\c{c}\~{a}o (Brazil) and Ministerio de Ciencia, Tecnolog\'{i}a e Innovaci\'{o}n Productiva (Argentina). We are grateful to the staffs of Gemini, the United Kingdom Infrared Telescope, and the NASA Infrared Telescope Facility for their support. We acknowledge the contribution of Christopher P. Morong who extracted the table of stars with magnitude less than 8 from the GLIMPSE Catalogue. T.O. was supported by NSF grant AST 1109014. F.N. acknowledges financial support through Spanish grants ESP2015-65597-C4-1-R and ESP2017-86582-C4-1-R (MINECO/FEDER). We also thank the referee for helpful comments. 

\end{acknowledgements}


\begin{thebibliography}{}

\bibitem[Clark et al.(2018)]{cla18}Clark, J. S., Lohr, M. E., Patrick, L. R., et al. 2018, A\&A, 618, A2

\bibitem[Cotera et al(1999)]{cot99}Cotera, A., Simpson, J. P., Erickson, E. E., et al. 1999, ApJ, 510, 747

\bibitem[Dong et al.(2011)]{don11}Dong, H., Wang, Q. D., Cotera, A., et al. 2011, MNRAS, 417, 114

\bibitem[Figer et al.(1998)]{fig98}Figer, D. F., Najarro, F., Morris, M., et al. 1998, ApJ, 506, 384

\bibitem[Figer, McLean, \& Morris(1999)]{fig99}Figer, D. F., McLean, I. S., \& Morris, M. 1999, ApJ, 514, 202

\bibitem[Fritz et al.(2011)]{fri11}Fritz, T. K., Gillesen, S., Dodds-Eden, K., et al. 2011, ApJ, 737, 73

\bibitem[Geballe, Najarro, \& Figer(2000)]{geb00}Geballe, T. R., Najarro, F., \& Figer, D. F. 2000, ApJL, 530, L57

\bibitem[Geballe \& Oka(2010)]{geb10}Geballe, T. R. \& Oka, T. 2010, ApJL, 709, L70

\bibitem[Geballe et al.(2011)]{geb11}Geballe, T. R., Najarro, F., Figer,ÊD.ÊF., et al.  2011, Nature, 479, 200

\bibitem[Glass et al.(19990)]{gla90}Glass, I. S., Moneti, A., \& Moorwood, A. F. M. 1990, MNRAS, 242, 55P

\bibitem[Goto et al.(2002)]{got02}Goto, M., McCall, B. J., Geballe, T. R., et al. 2002, PASJ, 54, 951

\bibitem[Goto et al(2011)]{got11}Goto, M., Usuda, T., Geballe, T. R., et al. 2011, PASJ, 63, L13

\bibitem[Indebetouw et al.(2005)]{ind05}Indebetouw, R., Mathis, J. S., Babler, B. L., et al. 2005, ApJ, 619, 931

\bibitem[Koyama et al.(1989)]{koy89}Koyama, K., Awaki, H., Kunied, S., et al. 1989, Nature, 339, 603

\bibitem[Lazio \& Cordes(1998)]{laz98}Lazio, T. J. W. \& Cordes, J. M. 1998, ApJ, 505, 715

\bibitem[Liermann, Hamann, \& Oskinova(2009)]{lie09}Liermann, A., Hamann, W.-R., \& Oskinova, L. M. 2009, A\&A, 494, 1137 

\bibitem[Mauerhan et al.(2010a)]{mau10a}Mauerhan, J. C., Muno, M. P., Morris, M. R., et al. 2010ba ApJ, 710, 706

\bibitem[Mauerhan et al.(2010b)]{mau10b}Mauerhan, J. R. Morris, M. R., Cotera, A., et al. 2010b, ApJ, 713, L33

\bibitem[Mauerhan et al.(2010c)]{mau10c}Mauerhan, J. C., Cotera, A., Dong, H., et al. 2010c, ApJ, 725, 188

\bibitem[Morris \& Serabyn(1996)]{mor96}Morris, M. \& Serabyn, E. 1996, ARA\&A, 34, 65

\bibitem[Muno et al.(2006)]{mun06}Muno, M. P., Bauer, F. E., Bandyopadhay, R. M., \& Wang, Q. D. 2006, ApJS, 165, 173

\bibitem[Nagata et al.(1990)]{nag90}Nagata, T., Woodward, C. E., Shure, M., Pipher, J. L., \& Okuda, H. 1990, ApJ, 351, 83

\bibitem[Nagata et al.(1993)]{nag93}Nagata, T, Hyland, A. R., Staw, S. M., Sato, S., \& Kawara, K. 1993, ApJ, 406, 501

\bibitem[Najarro et al.(2017)]{naj17}Najarro, F., Geballe, T. R., Figer, D. F., \& de la Fuente, D. 2017 ApJ, 845, 127 

\bibitem[Oka et al.(1998)]{oka98}Oka, T., Hasegawa, T., Hayashi, M., et al. 1998, ApJ, 493, 730

\bibitem[Oka et al.(2005)]{oka05}Oka, T., Geballe, T. R., Goto, M., et al. 2005, ApJ, 632, 882

\bibitem[Oka et al.(2019)]{oka19}Oka, T., Geballe, T. R., Goto, M., et al. 2019, in prep.

\bibitem[Okuda et al.(1990)]{oku90}Okuda, H., Shibai, H., Nakagawa, T., et al. 1990, ApJ, 351, 89

\bibitem[Ram\'irez et al.(2008)]{ram08}Ram\'irez, S. V., Arendt, R. G., Sellgren, K., et al. 2008, ApJS, 175, 147

\bibitem[Skrutskie et al.(2006)]{skr06}Skrutskie, M. F., Cutri, R. M., Stiening, M. D., et al. 2006, AJ, 131, 1166

\bibitem[Tuthill et al.(2006)]{tut06}Tuthill, P., Monnier, J., Tanner, A., et al. 2006, Sci, 313, 935

\bibitem[Yamauchi et al.(1990)]{yam90}Yamauchi, S., Kawada, K., Koyama, H., et al. 1990, ApJ, 365, 532 


\end{thebibliography}
\end{document}